%% file: gpu-graphit.tex
\documentclass[conference]{IEEEtran}
\usepackage{color}
\usepackage{xspace}

\usepackage{booktabs}   
\usepackage{subcaption} 
\usepackage{flushend}
\usepackage{graphicx}
\usepackage{balance}  
\usepackage{microtype}
\usepackage{soul}
\usepackage{booktabs}
\usepackage{subcaption}
\usepackage{xspace}
\usepackage[normalem]{ulem}
\usepackage{listings}%
\usepackage{xcolor}
\usepackage[font=small,labelfont=bf,aboveskip=2pt,belowskip=0pt]{caption}
\usepackage{parcolumns}
\usepackage{algorithm}
\usepackage[noend]{algpseudocode}
\usepackage{enumitem}

\usepackage{bm}
\usepackage{arydshln}
\usepackage{multirow}
\usepackage{tabu}

\input{graphitdef}

\newcommand{\graphit}{GraphIt\xspace}
\newcommand{\GG}{GG\xspace}
\newcommand{\GGfull}{GG GraphIt GPU compiler\xspace}
\newcommand{\gswitch}{GSWITCH\xspace}
\newcommand{\sgswitch}{GSW\xspace}
\newcommand{\sepgraph}{SEP-Graph\xspace}
\newcommand{\gunrock}{Gunrock\xspace}
\newcommand{\sgunrock}{GU\xspace}

\newcommand{\bfs}{BFS\xspace}
\newcommand{\cc}{CC\xspace}
\newcommand{\sssp}{SSSP\xspace}
\newcommand{\pr}{PR\xspace}
\newcommand{\bc}{BC\xspace}
\newcommand{\deltastepping}{Delta-Stepping\xspace}

\newcommand{\sgorkut}{OK\xspace}
\newcommand{\sgtwitter}{TW\xspace}
\newcommand{\sglivejournal}{LJ\xspace}
\newcommand{\sgsinaweibo}{SW\xspace}
\newcommand{\sghollywood}{HW\xspace}
\newcommand{\sgindochina}{IC\xspace}
\newcommand{\sgusad}{RU\xspace}
\newcommand{\sgroadca}{RN\xspace}
\newcommand{\sgroadcentral}{RC\xspace}
\usepackage{hyphenat}
\newcommand{\gorkut}{soc\hyp orkut\xspace}
\newcommand{\gtwitter}{soc\hyp twitter\hyp 2010\xspace}
\newcommand{\glivejournal}{soc\hyp LiveJournal\xspace}
\newcommand{\gsinaweibo}{soc\hyp sinaweibo\xspace}
\newcommand{\ghollywood}{hollywood\hyp 2009\xspace}
\newcommand{\gindochina}{indochina\hyp 2004\xspace}
\newcommand{\gusad}{road\_usa\xspace}
\newcommand{\groadca}{roadNet\hyp CA\xspace}
\newcommand{\groadcentral}{road\_central\xspace}

\newcommand{\TwceFullname}{Edge-based Thread Warps CTAs\xspace}

\newcommand{\twce}{ETWC\xspace}
\newcommand{\twc}{TWC\xspace}

\newcommand{\wm}{WM\xspace}
\newcommand{\cm}{CM\xspace}
\newcommand{\strict}{STRICT\xspace}
\newcommand{\vp}{VP\xspace}

\newcommand{\edgeblocking}{EdgeBlocking\xspace}
\newcommand{\cuda}{CUDA\xspace}
\newcommand{\kernelfusion}{kernel fusion\xspace}

\newcommand{\factorsbroad}{load balancing, traversal direction, active vertexset management, and work-efficiency\xspace}

\newcommand{\tablelst}[1]{{\lstinline[]$#1$}}

\newcommand{\stt}[1]{{\small \texttt{#1}}}

\newcommand{\OG}{PriorityGraph\xspace}

\newcommand{\uptospeedup}{5.11$\times$\xspace}

\newcommand{\fastercount}{66\xspace}

\newcommand{\totalexpr}{90\xspace}
\newcommand{\totalalgo}{5\xspace}
\newcommand{\totalgraph}{9\xspace}

\newenvironment{denseitemize}{
\begin{itemize} [topsep=2pt, partopsep=0pt, leftmargin=1.5em]
 \setlength{\topsep}{0pt}
 \setlength{\itemsep}{2pt}
 \setlength{\parskip}{0pt}
 \setlength{\parsep}{0pt}
}{\end{itemize}}

\newcolumntype{L}[1]{>{\raggedright\let\newline\\\arraybackslash\hspace{0pt}}p{#1}}
\newcolumntype{C}[1]{>{\centering\let\newline\\\arraybackslash\hspace{0pt}}p{#1}}
\newcolumntype{R}[1]{>{\raggedleft\let\newline\\\arraybackslash\hspace{0pt}}p{#1}}

\newcommand{\myparagraph}[1]{\noindent {\bf #1.}}

\usepackage{microtype}
\makeatletter
\newcommand{\linebreakand}{%
  \end{@IEEEauthorhalign}
  \hfill\mbox{}\par
  \mbox{}\hfill\begin{@IEEEauthorhalign}
}
\makeatother

\author {
	\IEEEauthorblockN {Ajay Brahmakshatriya}
	\IEEEauthorblockA {MIT CSAIL\\
			   USA\\
			   ajaybr@mit.edu}
\and 
	\IEEEauthorblockN {Yunming Zhang}
	\IEEEauthorblockA {MIT CSAIL\\
			   USA\\
			   yunming@mit.edu}
\and 
	\IEEEauthorblockN {Changwan Hong}
	\IEEEauthorblockA {MIT CSAIL\\
			   USA\\
			   changwan@mit.edu}
\linebreakand
	\IEEEauthorblockN {Shoaib Kamil}
	\IEEEauthorblockA {Adobe Research\\
			   USA\\
			   kamil@adobe.com}
\and
	\IEEEauthorblockN {Julian Shun}
	\IEEEauthorblockA {MIT CSAIL\\
			   USA\\
			   jshun@mit.edu}
\and
	\IEEEauthorblockN {Saman Amarasinghe}
	\IEEEauthorblockA {MIT CSAIL\\
			   USA\\
			   saman@csail.mit.edu}
}

\begin{document}

\title {Compilation Techniques for Graph Algorithms on GPUs}

\maketitle
\input{abstract}
\input{intro}

\input{relatedwork}

\input{optimizations}

\input{programmingmodel}

\input{schedulinglanguage}

\input{compilerimplementation}
\input{evaluation}
\input{conclusion}
\input{acks}

\newpage
\bibliographystyle{IEEEtran}

\bibliography{gpu-graphit.bib}
\end{document}

%% file: graphitdef.tex
\def\OPTL{\textrm{$[$}}
\def\OPTR{\textrm{$]$}}

\definecolor{lightbackground}{rgb}{.98,.98,.97}
\definecolor{darkgray}{rgb}{.3,.3,.3}
\definecolor{darkred}{rgb}{.6,0,0}
\definecolor{darkgreen}{rgb}{0,.6,0}
\definecolor{darkblue}{rgb}{0,0,.6}

\lstdefinelanguage{graphit}{%
  keywords={[1]},
  keywords={[2]func,end,element,const,var, field,for,%
    vertexset,edgeset,vector,priority_queue,%
    void,char,short,long,int,float,double,boolean,size_t, int32_t, delete,bool},
  keywords={[3]filter, from, to, srcFilter, dstFilter, apply, applyModified, applyUpdatePriority},
  keywords={[4],schedule,s0, s1 , s2, SimpleGPUSchedule, HybridGPUSchedule
    },
  literate={[OPT[}{{\OPTL}}1 {]OPT]}{{\OPTR}}1,
  string=[b]",
  comment=[l]//,
  morecomment=[s]{/*}{*/},
  mathescape=true,
  flexiblecolumns=true,
  tabsize=2,
  captionpos=b,
  frame=single,
  framerule=0pt,
  aboveskip=2pt,
  belowskip=1pt,
  framesep=1pt,
  basicstyle=\scriptsize\ttfamily,
  keywordstyle={[1]\color{darkred}},
  keywordstyle={[2]\color{blue}},
  keywordstyle={[3]\color{black}},
  keywordstyle={[4]\color{darkred}},
  numbers=left,
  stepnumber=1,
  numbersep=3pt,
  numberstyle=\tiny,
  emphstyle=\slshape,
  commentstyle=\color{darkgray},
  stringstyle=\color{darkgreen},
  xleftmargin=4.0ex,
}

\lstdefinelanguage{default}{
  basicstyle=\scriptsize\ttfamily,
}

\newcommand{\lstreset}{\lstset{language=default,backgroundcolor=\color{lightbackground},%
  belowskip=.5em, aboveskip=.5em}}

\lstreset

%% file: abstract.tex
\begin{abstract}
The performance of graph programs depends highly on the algorithm, the
size and structure of the input graphs, as well as the features of
the underlying hardware.  No single set of optimizations or one
hardware platform works well across all settings. To achieve high
performance, the programmer must carefully select which set of
optimizations and hardware platforms to use. The \graphit programming
language makes it easy for the programmer to write the algorithm once and optimize it for different inputs using a scheduling language. However, \graphit currently
has no support for generating high-performance code for GPUs.
Programmers must resort to re-implementing the entire algorithm from scratch in a low-level language with an entirely different set of abstractions and optimizations in order to achieve high performance on GPUs. 

We propose \GG, an extension to the \graphit compiler framework, that
achieves high performance on both CPUs and GPUs using the same
algorithm specification.  \GG significantly expands the optimization
space of GPU graph processing frameworks with a novel GPU scheduling
language and compiler that enables combining load balancing, edge
traversal direction, active vertexset creation, active vertexset
processing ordering, and kernel fusion optimizations.  \GG also
introduces two performance optimizations, \TwceFullname load
balancing (\twce) and \edgeblocking, to expand the optimization space
for GPUs.  \twce improves load balancing by dynamically partitioning the
edges of each vertex into blocks that are assigned to threads, warps,
and CTAs for execution.  \edgeblocking improves the locality of the
program by reordering the edges and restricting random memory accesses
to fit within the L2 cache.  We evaluate \GG on \totalalgo algorithms and
\totalgraph input graphs on both Pascal and Volta generation NVIDIA
GPUs, and show that it achieves up to \uptospeedup speedup over
state-of-the-art GPU graph processing frameworks, and is the fastest
on \fastercount out of the \totalexpr experiments.
\end{abstract}


%% file: intro.tex
\section{Introduction}
\label{sec:intro}

Graph processing is at the heart of many modern applications, such as
recommendation engines~\cite{Eksombatchai2018, pinsage2018}, social
network analytics~\cite{Sharma2016,TAOFacebook2012}, and map
services~\cite{Pallottino1998}. Achieving high performance is
important because these applications often need to process very large
graphs
and/or have strict latency
requirements~\cite{Eksombatchai2018}.

In prior work, we built the \graphit DSL compiler~\cite{graphit:2018, zhang2019prioritygraph} to generate high-performance CPU code from a high-level algorithm language. \graphit achieves state-of-the-art 
performance on CPUs across different algorithms and graph inputs by introducing a scheduling language
to tune optimizations. The algorithm language has primitives for topology-driven algorithms, data-driven algorithms, and priority-based algorithms. This algorithm and schedule representation makes it easy for the programmer to write an algorithm once and generate different highly-optimized implementations by simply changing the schedule. We will discuss in detail the \graphit algorithm and scheduling languages with examples in Section~\ref{sec:pmodel}.

\begin{figure}[!t]
    \centering
    \includegraphics [width=0.7\columnwidth] {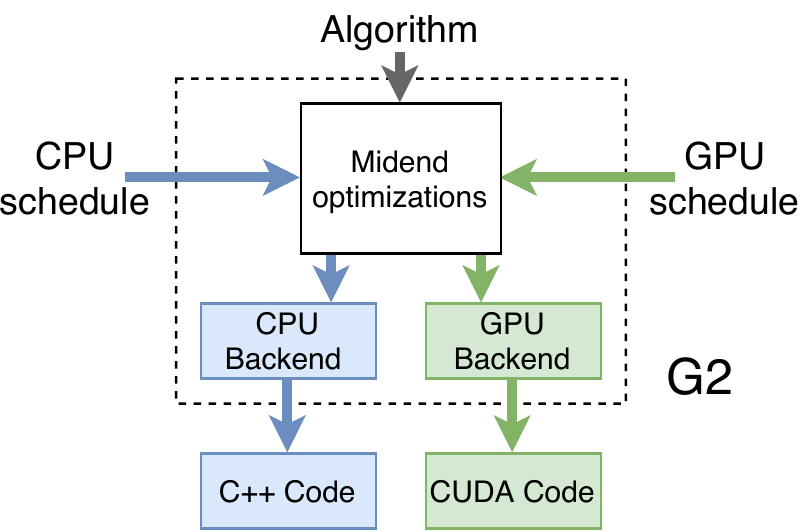}
    \caption{\GG adds a new GPU backend to \graphit that produces CUDA code from the same \graphit algorithm language and a scheduling language tailored for GPU optimizations. }
    \label{fig:3waydecouple}
\end{figure}

Apart from a large body of work for optimizing algorithms on different inputs on CPUs~\cite{graphit:2018,Sun2017,shun13ppopp-ligra,shun15dcc-ligraplus,Grossman2018,sundaram15vldb-graphmat,peng2018graphphi,Yunming2017,nguyen13sosp-galois}, researchers have also used different 
hardware platforms for high-performance graph processing, including GPUs~\cite{merrill2012scalable,Pai2016,ben2017groute,GSwitch2019,wang2019sep,liu2019simd,khorasani2014cusha,KhorasaniGPU2015,wang2017gunrock,hong2017multigraph},
and we have found that no single hardware is suitable for all
algorithms. Some algorithms perform better on CPUs and others perform
better on GPUs. Shared-memory CPUs have out-of-order execution, larger
caches, and larger memories than GPUs, which enables larger graphs to
be processed. At the same time, GPUs typically have an order of
magnitude more computing power and memory bandwidth~\cite{cuda2019},
and can better exploit data parallelism.

Unfortunately, since \graphit does not support code generation for GPUs, programmers have to resort to writing
CUDA implementations from scratch while building completely new abstractions and optimizations which rarely can be applied to different algorithms. 
Therefore, it is critical to build a common framework that can use a single algorithm representation to generate both CPU and GPU code, and
that also supports compiler transformations and compiler analyses specific to their respective platforms. 

We introduce \GG, an extension to the \graphit DSL compiler to generate high-performance
implementations on both CPUs and GPUs from the same high-level graph
algorithm specification. \graphit already demonstrated that
separating algorithm specification from scheduling representation is a
powerful way to optimize graph algorithms. We apply the same idea to GPUs by using a scheduling language to choose optimizations tailored specifically for GPUs. Figure~\ref{fig:3waydecouple} shows the new architecture of the \graphit compiler with the \GG extensions. 
Generating high-performance GPU
implementations presents significantly more challenges.  GPUs incur a
huge cost when accessing global memory in an uncoalesced fashion across
threads in a warp.  We have to optimize not just for temporal locality,
but also for locality across threads in a warp.  Achieving load
balancing is much harder on GPUs than on CPUs due to a large number of
execution units.  The GPU implementations must also handle data
transfer between the host and the device depending on what portion of the
code executes on the GPU.

Numerous performance optimizations have been proposed to address the
challenges of using GPUs for graph
processing~\cite{merrill2012scalable, nodehi2018tigr,
  10.1145/3342195.3387537}.  However, no existing GPU graph processing
framework supports all of the necessary optimizations to achieve high
performance, as shown in Table~\ref{tab:opt_space}.  GPU graph
processing libraries~\cite{wang2017gunrock, GSwitch2019, wang2019sep}
cannot easily support optimizations that require global program and
data structure transformations.  A compiler approach that employs
program transformations and code generation is more flexible.
However, existing compiler approaches~\cite{Pai2016} lack analysis
passes, code generation, and runtime library support for optimizations
such as direction-optimization and active vertexset creation.

\GG proposes a novel scheduling language, new program analyses, and a
code generation algorithm to support the wide variety of performance
optimizations while maintaining a simple high-level algorithm
specification.  The GPU scheduling language enables programmers to apply
load balancing, edge traversal direction, active vertexset creation,
active vertexset processing ordering, and kernel fusion optimizations.
The compiler uses a liveness analysis to reuse allocations for some
variables to improve memory efficiency.  \GG performs program analyses
and transformations on user-defined functions (UDFs) and iterative
edge traversals to incorporate optimizations that require global
program transformations, such as direction-optimization and
\kernelfusion. Figure~\ref{lst:bfs_gpu_schedule} the GPU scheduling language 
input for the same \bfs algorithm shown in Figure~\ref{lst:bfs_example}.

We also present two performance optimizations, \edgeblocking and
\TwceFullname (\twce), to further expand the space of graph
optimizations on GPUs.  \edgeblocking reorders the edges in coordinate
format (COO) to group together edges accessing segments of vertices
that fit in the L2 cache of the GPU.  This optimization improves
performance by keeping random memory accesses within the fast L2
cache, reducing slower global memory accesses.  \twce partitions the
edges of active vertices into small, medium, and large buckets that
can be processed by groups of threads, warps, and Cooperative Thread
Arrays (CTAs), respectively, to improve load balancing and utilization
of Streaming Multiprocessors (SMs).

We evaluate the performance of \GG on five graph algorithms: PageRank
(\pr), Breadth-First Search (\bfs), Delta-Stepping for Single-Source
Shortest Paths (\sssp), Connected Components (\cc), and Betweenness
Centrality (\bc). \GG is built as an extension to
\graphit~\cite{graphit:2018, zhang2019prioritygraph}.  We are able to
obtain up to \uptospeedup speedup over the fastest of the
state-of-the-art GPU graph frameworks Gunrock~\cite{wang2017gunrock},
GSwitch~\cite{GSwitch2019}, and SEP-graph~\cite{wang2019sep}.  \GG
generates CPU implementations that match the performance of the
original \graphit compiler, thus not compromising CPU performance.

This paper makes the following contributions:
\begin{denseitemize}
\item An analysis of the fundamental tradeoffs among locality,
  work-efficiency, and parallelism in GPU graph optimizations
  (Section~\ref{sec:optimizations}).
\item A novel GPU scheduling language that allows programmers to
  combine load balancing, edge traversal direction, active vertexset
  creation, active vertexset processing ordering, and kernel fusion
  optimizations (Section~\ref{sec:scheduling}).
\item A compiler with program analyses and transformations to support
  efficient code generation on both CPU and GPU platforms
  (Sections~\ref{sec:pmodel} and~\ref{sec:compiler}).
\item Two performance optimizations for GPUs--- \edgeblocking,
  which improves the locality of edge processing, and \twce, which
  improves load balancing (Section~\ref{sec:optimizations}).
\item A comprehensive evaluation of \GG on GPUs showing that \GG
  outperforms state-of-the-art GPU graph processing frameworks by up
  to \uptospeedup on \totalexpr experiments across two generations of
  NVIDIA GPUs (Section~\ref{sec:evaluation}).
\end{denseitemize}

%% file: relatedwork.tex
\vspace{-0.5em}
\section{Related Work}
\label{sec:related}

\begin{table*} [!th]
\vspace{2mm}
\centering
\label{tab:opt_space}
\scriptsize{
\begin{tabu} {|>{\raggedright}p{3cm}|>{\raggedright}p{2.5cm}|>{\raggedright}p{1.5cm}|>{\raggedright}p{1.5cm}|>{\raggedright}p{2.5cm}|>{\raggedright}p{3.0cm}|}
\hline
Framework & \textbf{\GG} & \gswitch & \sepgraph & \gunrock & IrGL \\ \hline
Load Balancing & WM, CM, ETWC, TWC, Strict, VertexBased, EdgeOnly & CM, WM, Strict, TWC & CM & VertexBased, TWC, EdgeBased & Serial, TB, Warp, FineGrained, TB+Warp, TB+FG, WP+FG, TB+Warp+FG \\ \hline
Edge Blocking & Supported & Not-Supported & Not-Supported & Not-Supported & Not-Supported \\ \hline
Vertexset Creation & SparseQueue, Bitmap, Boolmap & SparseQueue, Bitmap & SparseQueue & SparseQueue, Bitmap & SparseQueue \\ \hline
Kernel Fusion & Supported & Not-Supported & Supported & Not-Supported & Supported \\ \hline
Direction-Optimization & Supported & Supported & Supported & Supported & Not-Supported \\ \hline
Vertexset Deduplication & Supported & Not-Supported & Supported & Supported & Not-Supported\\ \hline
Vertices Ordering & Supported & Supported & Supported & Supported & Supported\\ \hline
Total Opt. Combinations & \textbf{576} & 32 & 16 & 48 & 32 \\\hline
\end{tabu}
}
\caption{Number of available optimizations in state-of-the-art GPU graph processing frameworks. Total is the product of all of the options.}
\label{tab:opt_space}
\vspace{-2em}
\end{table*}

\myparagraph{Compilers for Graph Programs}
\GG is built as
an extension to our previous work \graphit~\cite{graphit:2018, zhang2019prioritygraph},
which is a domain-specific language and compiler that expands the
optimization space to outperform other CPU frameworks by decoupling
algorithms and optimizations. Apart from adding support for GPUs, \GG
also adds new mid-end optimization passes, which can potentially be
reused for other architectures. \GG  creates a novel GPU
scheduling language that enables users to explore a much wider
space of optimizations for GPUs.

IrGL~\cite{Pai2016} is another compiler framework that introduces a
programming language and an intermediate representation specifically
for graph applications on GPUs. IrGL first proposed the kernel fusion
optimization, which reduces the overhead of kernel launch for \sssp
and \bfs on high-diameter road networks. However, IrGL supports a much
narrower space of optimizations (see
Table~\ref{tab:opt_space}). Unlike \GG, IrGL does not
support code generation for CPUs and GPUs from the same high-level
representation.

\myparagraph{Graph Frameworks for both CPUs and GPUs}
GraphBLAS~\cite{kepner2016mathematical, YangGraphBlas2018} uses sparse
linear algebra operations as building blocks for graph algorithms.
The GraphBLAS approach does not support ordered algorithms, such as
Delta-Stepping for SSSP~\cite{MEYER2003114}, which are needed for achieving high
performance on SSSP and other problems~\cite{GSwitch2019,
  wang2017gunrock, wang2019sep}.  \GG supports ordered graph
algorithms with a GPU-based two-bucket priority queue.
Abelian~\cite{abelian2018} uses the Galois framework as an interface
for shared-memory CPU, distributed-memory CPU, and GPU platforms.
However, it lacks support for direction-optimization, various load
balancing optimizations, and active vertexset creation optimizations,
which are needed to achieve high performance.

\myparagraph{Graph Frameworks for GPUs} There has been a tremendous amount of effort on developing high-performance graph processing frameworks on
GPUs
(e.g.,~\cite{merrill2012scalable,Pai2016,ben2017groute,wang2019sep,liu2019simd,nodehi2018tigr,khorasani2014cusha,KhorasaniGPU2015,wang2017gunrock,harish2007accelerating,hong2011accelerating,liu2015enterprise,shi2017frog,hong2017multigraph,davidson2014work,soman2010fast,nasre2013data,che2014gascl,kim2016gts,gaihre2019xbfs,han2017graphie}). Gunrock~\cite{wang2017gunrock}
proposes a novel data-centric abstraction and incorporates existing
GPU optimization strategies. GSWITCH~\cite{GSwitch2019} identifies
and implements a set of useful optimizations and uses auto-tuning to
achieve high performance. DiGraph~\cite{10.1145/3297858.3304029} is another framework that targets multiple GPUs
and achieves higher throughput by employing a novel path scheduling strategy that minimizes communication and makes the algorithms converge in fewer iterations. 
However, most of the frameworks support only a subset
of existing optimizations and cannot achieve high performance on all
algorithms, graphs, and GPU architectures
(see Table~\ref{tab:opt_space}).

\myparagraph{Other Graph Processing Frameworks} There has been a large
body of work on graph processing for shared-memory multicores
(e.g.,~\cite{Sun2017,shun13ppopp-ligra,shun15dcc-ligraplus,zhang15ppopp-numa-polymer,Grossman2018,sundaram15vldb-graphmat,peng2018graphphi,
  Hong12asplos,socialite13ICDE,abelian2018,Aberger2016,KickStarter2017,Yunming2017}),
distributed memory
(e.g.,~\cite{Chen15PowerLyra,Zhu16gemni,gluon2018,dathathri2019gluon,McCune2015,conf/uai/LowGKBGH10,Gonzalez2012,malewicz10sigmod-pregel,prabhakaran12atc-grace}),
and external memory
(e.g.,~\cite{kyrola12osdi-graphchi,roy13sosp-xstream,Vora2016,Wang15ATC,zhang2018wonderland,zuo2019grapple,Maass2017,Graspan2017,Zhu15ATC-GridGraph,zheng15fast-flashgraph,wang2018rstream}). These
frameworks support a limited set of optimizations and cannot achieve
consistent high performance across different algorithms and
graphs~\cite{graphit:2018}.  

\myparagraph{Load Balancing Optimizations on GPU} \twc is a dynamic
load balancing strategy and prefix sum-based frontier computation
designed for efficient breadth-first search on
GPUs~\cite{merrill2012scalable}. Gunrock~\cite{wang2017gunrock} and
GSwitch~\cite{GSwitch2019} implement both \twc and a few other static
load balancing techniques, which we describe in more detail in
Section~\ref{sec:optimizations}.  Tigr~\cite{nodehi2018tigr} and
Subway~\cite{10.1145/3342195.3387537} achieve load balancing by
preprocessing the graph to ensure that all the vertices have a similar number
of neighbors. This approach incurs significant prepossessing overhead
and does not generalize across all graph algorithms. Tigr also exposes a
much smaller optimization space and does not show how the load
balancing techniques can be combined with other optimizations. \GG
introduces the \twce scheme, which reduces the runtime overhead in
exchange for some load balance.

\myparagraph{Locality-Enhancing Optimizations} The irregular memory
access pattern in graph algorithms makes it hard to take advantage of
the memory hierarchy on CPUs and GPUs~\cite{Beamer15IISWC,
  Yunming2017}. On CPUs, locality can be enhanced by tiling the
graph~\cite{Yunming2017,zhang15ppopp-numa-polymer,Zhu16gemni,Sun2017}
or by reordering the memory accesses at runtime~\cite{Beamer2017,
  Kiriansky2016}. Reordering the memory accesses at runtime on GPUs is
not practical because there is not enough memory to buffer all of the
irregular memory updates and the performance improvement cannot compensate for the extra work to be done. A naive graph tiling approach on GPUs results
in poor performance due to insufficient parallelism. \edgeblocking
finds a good balance between locality and parallelism by tiling for
the shared L2 cache and processing one tiled subgraph at a time.
Previous work~\cite{nagasaka14icpads-sparse, HongTiling2019} has
used tiling in sparse matrix computations.

%% file: optimizations.tex
\section{GPU Optimization Space} 
\label{sec:optimizations}

Optimizations that make tradeoffs among locality, parallelism, and
work-efficiency (number of instructions) can improve the performance of
graphs algorithms by orders of magnitude over naive
implementations~\cite{merrill2012scalable, wang2017gunrock,
  GSwitch2019} on GPUs.  In this section, we describe the
tradeoff space of existing graph optimizations for GPUs.

\myparagraph{Active Vertexset Creation} 
  Different ways of creating
output frontiers have different tradeoffs. These creation mechanisms
include fused~\cite{GSwitch2019} (with the edge traversal) vs. unfused
and different representations such as bitmaps, bytemaps, and sparse
arrays. The fused mode trades off parallelism for better
work-efficiency. Bitmaps have better locality but suffer from atomic
synchronization overhead, unlike bytemaps.

\myparagraph{Kernel Fusion across Iterations} The iterative nature of many
graph algorithms can be expressed with a while loop. A GPU kernel can
be launched in each iteration in the while loop, but if there is not
enough work per iteration, the kernel launch overhead can dominate the
running time. To alleviate this overhead, the while loop can be moved
into the kernel so that the GPU kernel is launched only
once~\cite{Pai2016}. This improves work-efficiency but sacrifices
parallelism and load balancing.

\myparagraph{Direction-Optimization}
The push (each vertex updates its neighbors) and pull (each destination reads its neighbors and updates itself) modes for updating vertex data offer a tradeoff between work-efficiency, atomic synchronization overheads, and parallelism. Direction-optimization achieves the best of both worlds by dynamically switching between the two directions based on the frontier size~\cite{Beamer-2012,shun13ppopp-ligra,Besta2017}.

\myparagraph{Active Vertexset Deduplication} Since active vertices may
share common neighbors, some vertices can appear multiple times in the
next frontier, which may cause redundant computation and even lead to
incorrect results. This can be avoided by an explicit deduplication
step, which affects work-efficiency, increasing or decreasing it
depending on the number of duplicate vertices.

\myparagraph{Active Vertexset Processing Ordering} The active vertices
can be processed according to a priority-based ordering, which can
significantly improve the work-efficiency for ordered graph algorithms
such as Delta-Stepping for SSSP, but at the expense of reducing
parallelism~\cite{MEYER2003114, zhang2019prioritygraph, GSwitch2019}.

\myparagraph{Load Balancing} Different load balancing schemes have
tradeoffs between parallelism and work-efficiency to various
degrees. Warp Mapping (\wm) and CTA Mapping
(\cm)~\cite{GSwitch2019,KhorasaniGPU2015} divide the active vertices
evenly across different warps and CTAs, and achieve high parallelism
but sacrifice work-efficiency due to overheads in partitioning the
active vertices.  \strict incurs even higher overheads but ensures
that each thread in the GPU processes the same number of edges. \twc
splits active vertices into buckets processed by threads, warps, or
CTAs based on their degrees~\cite{merrill2012scalable,
  wang2017gunrock}. \twc has smaller runtime overhead (high
work-efficiency), but a lower degree of parallelism compared to \wm,
\cm, and \strict. However, for some graphs, this overhead can still be large compared to the speedup achieved with better load balancing. Vertex-Parallel (\vp) simply maps each vertex to a
thread and has the least overhead but poor load balancing.

In this paper, we introduce two optimizations: \twce, a load balancing strategy that has reduced runtime overhead as compared to \twc; and \edgeblocking, a graph partitioning optimization for improving locality of vertex data accesses. Details of these optimizations are described in Section~\ref{sec:compiler}.

%% file: programmingmodel.tex
\section{Algorithm Language} 
\label{sec:pmodel}

\begin{figure} [t]
\vspace{2mm}
\begin{lstlisting} [language=graphit,escapechar=|]
element Vertex end |\label{line:globalDeclStart} |
element Edge end
const edges : edgeset{Edge}(Vertex,Vertex) 
    = load(argv[1]);
const vertices : vertexset{Vertex} 
    = edges.getVertices();
const parent : vector{Vertex}(int) = -1; |\label{line:globalDeclEnd}|

func toFilter(v : Vertex) -> output : bool
	output =  (parent[v] == -1); |\label{line:parent_filter}|
end

func updateEdge(src : Vertex, dst : Vertex)
	parent[dst] = src; |\label{line:parent_update}|
end

func main()
	var frontier : vertexset{Vertex} |\label{line:frontier_init}|
	    = new vertexset{Vertex}(0); 
	var start_vertex : int = atoi(argv[2]);
	frontier.addVertex(start_vertex);
	parent[start_vertex] = start_vertex;
	#s0# while (frontier.getVertexSetSize() != 0) |\label{line:labels0}|
		#s1# var output : vertexset{Vertex} = 
			edges.from(frontier).to(toFilter).
				applyModified(updateEdge, parent, true); |\label{line:applymodified}|
		delete frontier; |\label{line:swap_start}|
		frontier = output;  |\label{line:swap_end}|
	end
	delete frontier;
end
\end{lstlisting}
\caption{\bfs program written in the \graphit algorithm language.}

\label{lst:bfs_example}
\end{figure}

\begin{figure}[!tbh]
\begin{lstlisting}[language=graphit,escapechar=|]
configApplyDirection("s0:s1", "DensePull-SparsePush")
configApplyDenseVertexSet("s0:s1", "src-vertexset", 
    "bitvector", "DensePull")
configApplyParallelization("s0:s1", 
    "dynamic-vertex-parallel") 
\end{lstlisting}
\caption{\graphit scheduling language input for optimizing the \bfs algorithm in Figure~\ref{lst:bfs_example} on CPUs using a hybrid traversal}
\label{lst:bfs_cpu_schedule}
\end{figure}

\begin{figure}[!tbh]
\begin{lstlisting}[language=graphit,escapechar=|]
SimpleGPUSchedule s1;
s1.configDirection(PUSH);
s1.configLoadBalance(VERTEX_BASED);
SimpleGPUSchedule s2 = s1;
s2.configDirection(PULL, BITMAP);
s2.configDeduplication(DISABLED);
s2.configLoadBalance(VERTEX_BASED);
s2.configFrontierCreation(UNFUSED_BITMAP);
HybridGPUSchedule h1(VERTEXSET_SIZE, "argv[3]", s1, s2); 
program->applyGPUSchedule("s0:s1", h1);
\end{lstlisting}
\caption{\GG scheduling language input for optimizing the \bfs algorithm in Figure~\ref{lst:bfs_example} on GPUs using a hybrid traversal. Since GPUs requires completely different transformations, the optimizations enabled are different from those in Figure~\ref{lst:bfs_cpu_schedule}.}
\label{lst:bfs_gpu_schedule}
\end{figure}

\GG separates the algorithm logic from performance optimizations with
a separate algorithm language and scheduling language. Since the goal of \GG is to generate 
high-performance GPU code with minimum effort while moving from CPUs, we use the same algorithm language from \graphit
which supports unordered graph algorithms~\cite{graphit:2018}, along with the extensions in \OG~\cite{zhang2019prioritygraph} to support ordered
graph algorithms. The algorithm language works with vectors, vertex and edge sets,
functional operators over sets, and priority queue operators, and
hides all of the low-level details needed for correctness and
performance, such as atomic synchronization and bit manipulation,
from the programmer.

Figure~\ref{lst:bfs_example} shows an example of the breadth-first
search (BFS) algorithm written in the \graphit algorithm language. The
program starts with a vertexset with just the
\stt{start\_vertex} and repeatedly updates the neighbors of the
vertices in the frontier by applying the \stt{updateEdge} and
\stt{toFilter} user-defined functions, while also creating the next
frontier from the vertices just updated. This is done till the frontier is empty.

Figure~\ref{lst:bfs_cpu_schedule} shows the scheduling language input
used in \graphit for optimizing this algorithm on CPUs with direction-optimization (hybrid traversal). The scheduling input enables hybrid
traversal with bitvector representation for the source vertex frontier
when traversing in the PULL direction. This schedule also enables
dynamic vertex-based parallelism. We have modified \graphit in such a
way that preserves the scheduling input for CPUs without significant
change in the generated code. Figure~\ref{lst:bfs_gpu_schedule} shows
the scheduling input used in \GG for the \bfs algorithm when
generating code for GPUs. This schedule also enables hybrid traversal
but also tunes various GPU specific scheduling options such as
deduplication, load balancing, and active vertexset creation strategy. These options will be explained in
detail in Section~\ref{sec:scheduling}.

%% file: schedulinglanguage.tex
\section{Scheduling Language}
\label{sec:scheduling}
We propose a novel scheduling language for GPUs that allows
users to combine the \factorsbroad optimizations described in
Section~\ref{sec:optimizations}. These
optimizations are different from the ones for CPUs because GPUs
have drastically
different hardware features, such as a larger number of threads, larger
memory bandwidth, smaller L1 and L2 cache per thread, smaller total
memory, and different synchronization overheads.

The scheduling language supports two main types of schedules,
\stt{SimpleGPUSchedule} and \stt{HybridGPUSchedule}. The
programmer can use them together to create complex schedules that are
best suited for their algorithm and graph type.

The \stt{SimpleGPUSchedule} object directly controls the
scheduling choices related to \factorsbroad. As shown in
Table~\ref{table:simplegpuschedule}, the objects have six
\stt{config*} functions. The programmer can use these functions
to specify the load balancing strategy, the edge traversal direction,
the output frontier creation strategy, whether deduplication is
enabled, the delta value for priority queues, and whether
\kernelfusion is applied to a particular loop along with some
optional parameters.

\begin{table*}[t]
\vspace{2mm}
\centering 
\begin {tabular} {p{3.5cm}|p{3cm}|p{10.5cm}}
Config function & Parameters & Description \\ \cline {1-3}
\tablelst{SimpleGPUSchedule} & \tablelst{SimpleGPUSchedule s0} \textbf{(optional)} & Create a new SimpleGPUSchedule object. Optional copy constructor argument.  \\ \cline{1-3}
\multicolumn{3}{l}{\textbf{Scheduling functions for {\lstinline[]$edges.appply$}}} \\ \cline{1-3}
\tablelst{configLoadBalance} & \tablelst{load_balance_type}, \tablelst{blocking_type} \textbf{(optional)}, \tablelst{int32_t blocking_size} & Configure the load balancing scheme to be used from one of \textbf{\tablelst{VERTEX_BASED}}, \tablelst{CM}, \tablelst{WM}, \tablelst{STRICT}, \tablelst{EDGE_ONLY}, \tablelst{ETWC}, and \tablelst{TWC}. Optionally enable blocking and set blocking size. \\ \cline{1-3}
\tablelst{configDirection} & \tablelst{direction}, \tablelst{front_rep} \textbf{(optional)}& Configure the direction of updates of vertex data between \textbf{\tablelst{PUSH}} and \tablelst{PULL}. Optionally also configure the representation of the frontier (\textbf{\tablelst{BOOLMAP}} or \tablelst{BITMAP}) when direction is \tablelst{PULL}.  \\ \cline{1-3}
\tablelst{configFrontierCreation} & \tablelst{frontier_type} & Configure how the output frontier will be created. Options are \textbf{\tablelst{FUSED}}, \tablelst{UNFUSED_BOOLMAP}, and \tablelst{UNFUSED_BITMAP}. \\ \cline{1-3}
\tablelst{configDeduplication} & \tablelst{dedup_enable}, \tablelst{dedup_type} & Configure if deduplication needs to be enabled when producing the output frontier (\textbf{\tablelst{ENABLED}}). The strategy can be \textbf{\tablelst{MONOTONIC_COUNTERS}}, \tablelst{BITMAP}, or \tablelst{BOOLMAP}. \\ \cline{1-3}
\multicolumn{3}{l}{\textbf{Scheduling functions for {\lstinline[]$edges.appplyUpdatePriority$}}} \\ \cline{1-3}
\tablelst{configDelta} & \tablelst{int32_t delta} & Configure the delta value to be used when creating the buckets in the priority queue. \\ \cline{1-3}
\multicolumn{3}{l}{\textbf{Scheduling functions for {\lstinline[]$while(condition)$}}} \\ \cline{1-3}
\tablelst{configKernelFusion} & \tablelst{enable_fusion} & Configure if kernel fusion is \textbf{\tablelst{ENABLED}} or \tablelst{DISABLED} for this while loop. \\ \cline{1-3}

\end{tabular}
\caption {Description of the SimpleGPUScheduling type and associated
  \lstinline{config} functions. The default value for each parameter
  is in \textbf{bold}.}
\label {table:simplegpuschedule}
\end{table*}

\begin {table*}[t]
\centering
\begin {tabular} {p{3cm}|p{3cm}|p{11cm}}
Function & Parameter & Desription \\ \cline{1-3}

\tablelst{HybridGPUSchedule} & \tablelst{hybrid_criteria}, \tablelst{float threshold}, \tablelst{SimpleGPUSchedule s1}, \tablelst{SimpleGPUSchedule s2} & Create a HybridGPUSchedule object by combining two SimpleGPUSchedule objects with a runtime condition (currently can be only \textbf{\tablelst{INPUT_VERTEXSET_SIZE}}) and a threshold. \\ \cline{1-3}
\end {tabular}
\caption {Description of the HybridGPUScheduling type and constructor. Default value for each parameter is shown in \textbf{bold}.}
\label {table:hybridgpuschedule}
\vspace{-1em}
\end {table*}

The \stt{HybridGPUSchedule} objects combine two \stt{SimpleGPUSchedule} objects with some runtime condition. The two \stt{SimpleGPUSchedule} objects can be entirely different, using different load balancing schemes, frontier creation types, traversal directions, etc. Depending on whether the runtime condition evaluates to true or false, one of the two \stt{SimpleGPUSchedule} objects is invoked. This API enables the programmer to implement complex optimizations like direction-optimization by combining two \stt{SimpleGPUSchedule} objects (Table~\ref{table:hybridgpuschedule}).

\begin{figure*}[h]
\includegraphics[width=\textwidth]{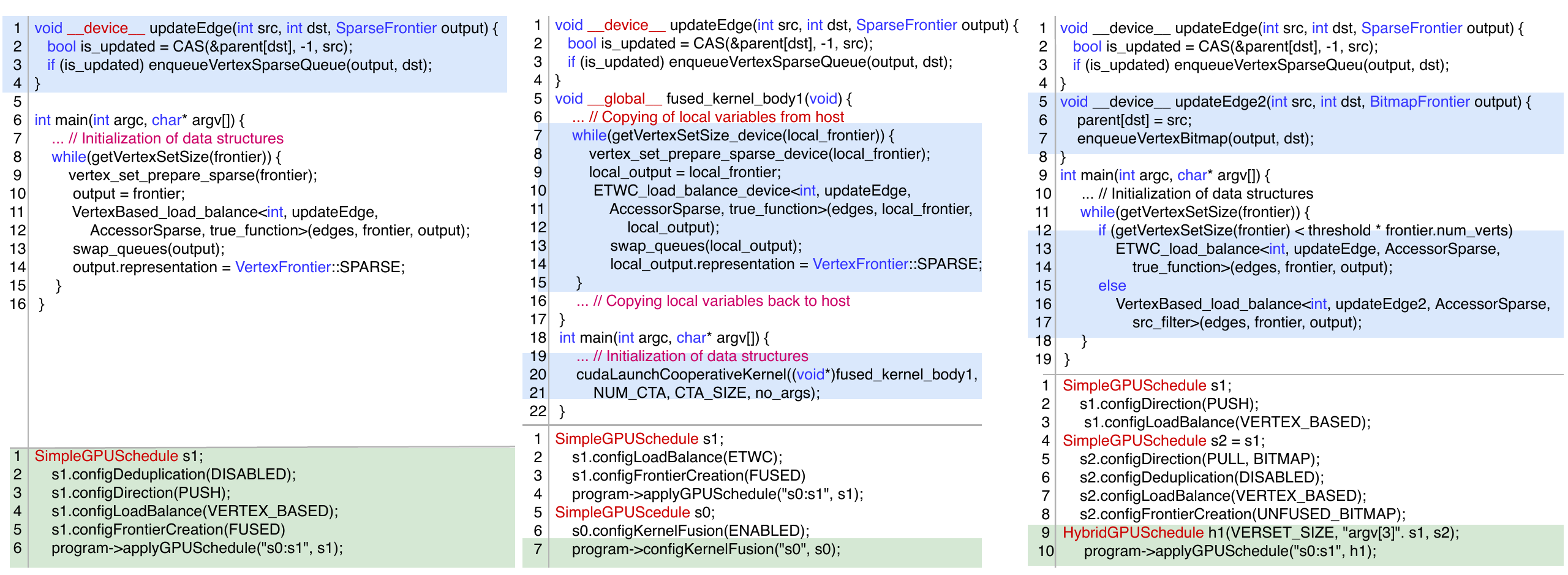}
\caption{Different schedules for a default \stt{PUSH}-based implementation (left), an implementation with \kernelfusion enabled (middle), and an implementation with direction-optimization (right). The code at the top shows snippets of the generated \cuda code for each of the schedules.}
\label{fig:3schedules}
\vspace{-1em}
\end{figure*}

The scheduling language shares some key features with the existing CPU
scheduling language, such as the ability to choose the edge traversal
direction, whether deduplication is enabled, and whether the iteration
is vertex-parallel or edge-parallel (based on load balancing type). These features are utilized by
the high-level compiler to perform target-independent
optimizations. 

\myparagraph{Scheduling for \bfs} Figure~\ref{fig:3schedules} shows
three different schedules that can be applied to the same \bfs
program. The first example shows the default \stt{PUSH} schedule
highlighted in green. Above the schedule, we show snippets of the corresponding
generated CUDA code from \GG. Notice that even with the default
schedule, the UDF (\stt{updateEdge}, highlighted in blue) has
been transformed to use an atomic compare-and-swap instruction when
updating the destination vertex data. This is because atomics are
required here for correctness, not for optimization.

The second example schedule shows how \kernelfusion can be applied to
an entire loop body. Notice that in the generated code, the entire
while loop is moved to the CUDA kernel and the main function just
calls this kernel. This schedule is suitable for high-diameter graphs
that have low parallelism within a round (e.g., road networks).

The third example shows a direction-optimizing schedule. We create two
individual \stt{SimpleGPUSchedule} objects, \stt{s1} and
\stt{s2}, and configure one for the \stt{PUSH} direction
and the other for the \stt{PULL} direction, along with other
parameters. We then combine the two into a single
\stt{HybridGPUSchedule} object that chooses between the two
based on the size of the input frontier and apply this to the
\stt{applyModified} operator. The generated code has a runtime
condition based on the size of the vertexset that chooses between two
separate implementations. Two versions of the UDF are created
because the edge traversal does not
require atomics when iterating in the \stt{PULL} direction. This schedule is suitable for power-law graphs (e.g., social network
graphs) that have significantly different numbers of active vertices in different
rounds.

%% file: compilerimplementation.tex
\section{Compiler Implementation}
\label{sec:compiler}

This section details the changes that \GG makes to the \graphit
compiler to support the new GPU backend. This includes analyses and
transformations added to the mid-end and the backend for generating
code for both the host and device using the parameters from the new
GPU scheduling language. We also describe the implementations of the
 \edgeblocking and \twce optimizations.

\subsection{Liveness Analysis for Frontier Reuse}
\GG adds a new liveness analysis to the mid-end to reduce costly memory
allocations on GPUs. The \stt{edges.applyModified} function
generally needs to allocate memory for the frontier that it
outputs, but if the frontier it takes as an input is not required
anymore, the allocated memory can be reused to avoid calls to the
costly \stt{cudaMalloc} and \stt{cudaFree} functions
(particularly useful when called inside a loop). To determine whether
the memory for frontiers can be reused, the compiler performs a
liveness analysis on the variables in the \stt{main}
function. If the analysis shows that the input frontier is deleted
before being used again, it sets the \stt{can\_reuse\_frontier} flag in
\stt{applyModified}.

\subsection{UDF Transformations}
\GG implements a dependence analysis pass to automatically insert atomic
instructions for updates to vertex data in UDFs.  Whether an
update requires atomics depends on which set of vertices are iterated
over in parallel by different threads, which in turn depends on the load
balancing strategy used. Therefore, the analysis pass needs to be made aware of the
different types of parallelism offered by the load balancing strategies
added by the GPU scheduling language. The analysis combines the results of
the dependence analysis and read/write analysis to identify updates
that are shared by multiple threads and inserts
atomics.

Apart from inserting atomics, the compiler also transforms the UDF to enqueue the vertices into the output frontier when their vertex data is updated. 
The compiler refers to the schedule and the tracking variable to select the enqueue function to use.

\subsection{GPU Backend Implementation}
\label{sec:gpu_backend_impl}
The GPU backend performs transformations and optimizations required
for generating CUDA code, including transfer of data between the host
and device, load balancing for a hierarchy of threads, inter-thread
and inter-warp communication, and \kernelfusion. The GPU backend
generates host code that takes care of loading graphs, allocating data
structures, and executing outer loops. It also generates device code that
actually implements the edge/vertexset iteration operators.

The GPU backend also implements the \kernelfusion optimization. As
shown on Line~\ref{line:labels0} of Figure~\ref{lst:bfs_example}, we
can annotate an entire loop to be fused using a label. The programmer
can then enable \kernelfusion with the
\stt{configKernelFunction} function. With this schedule, the
compiler must launch a single \cuda kernel for the entire loop. This
means that all steps inside the body of the loop must execute on the
device. \GG first performs an analysis to figure out if the body
contains a statement that it cannot execute on the device (e.g., creation of objects that require allocation of memory or a
call to the \stt{delete} operator, if the hardware/runtime does
not support dynamic memory allocation).

\GG generates code for the fused kernel by iterating over the program
AST multiple times. In the first pass, the backend generates a
\stt{\_\_global\_\_} kernel that has the actual while loop and calls
to the functions for the operators inside the loop. Here, the compiler
also  handles local variables referenced from the
\stt{main} function by hoisting them to \stt{\_\_device\_\_}
global variables. Usually, the operator requires a lot more threads than the ones launched for the single kernel. \GG inserts another 
\stt{for} loop around the operator code so that each thread can do work for more threads
serially. In the second pass, \GG actually generates a
call to the CUDA kernel where the \stt{while} loop is in the
\stt{main} function. It also copies all required local variables
to the newly-hoisted global variables before calling the kernel and
then copies them back after the kernel returns.

\myparagraph{Load Balancing Library} Effective load balancing is one of the key factors for obtaining high performance on GPUs. By ensuring that each thread has an almost equal number of edges to process, we minimize the time that threads are waiting for other threads to finish. Load balancing within a warp is even more critical to minimize divergence in some generations of GPUs where diverged threads are executed serially.
However, precise load balancing comes at a cost. Before actually processing the edges, threads have to coordinate to divide 
work among CTAs, warps, and threads, which involves synchronization, global memory accesses, and extra computations. 
Different load-balancing schemes provide a tradeoff between cost and the amount of balancing achieved. Which load balancing strategy
is the most beneficial depends on the algorithm and the graph input. As a result, \GG currently adds support for a total
of 7 load balancing strategies. A load-balancing library creates an abstraction with a 
templated interface that processes a set of edges after dividing them between threads. This modular design not only makes code generation easy but also makes it easy to add more load balancing techniques in the future. The core library has both \stt{\_\_global\_\_} and \stt{\_\_device\_\_} wrappers, allowing these routines to be reused inside the fused kernel 
with \kernelfusion.

\subsection{GPU-Specific Optimizations}
\label{sec:new-optimizations}
This section explains the two optimizations (\edgeblocking and
\twce) \GG adds to the \graphit compiler to boost performance on GPUs
by improving cache utilization and load balancing, respectively.

\myparagraph{\edgeblocking} We propose the new \edgeblocking
optimization, which tiles the edges into a series of subgraphs to
improve the locality of memory accesses.
Algorithm~\ref{algo:preprocessing} shows the steps for preprocessing
an input graph to apply the \edgeblocking optimization. The
preprocessing is a two-step process. The first for-loop
(Line~\ref{line:preprocessloop1}--8 of
Algorithm~\ref{algo:preprocessing}) iterates through all of the edges
and counts the number of edges in each subgraph. The algorithm then
uses a prefix sum on these counts to identify the starting point for
each subgraph in the output graph's edges array. The second for-loop
(Line~\ref{line:preprocessloop2}--14 of
Algorithm~\ref{algo:preprocessing}) then iterates over each edge again
and writes it to the appropriate subgraph while incrementing that
subgraph's counter.  We apply the function \stt{process\_edge} to
each edge as shown in Algorithm~\ref{algo:edgprocessing}. The
arguments to this function are the source vertex and the destination
vertex.  The pseudocode shown in Algorithm~\ref{algo:edgprocessing} is
executed by each thread in a thread block.  We use an outer for-loop
(Line~\ref{line:edgeprocessingouterloop} of
Algorithm~\ref{algo:edgprocessing}) that iterates over each
subgraph. Within each subgraph, the edges are then processed by all of
the threads using a \stt{cooperative for}
(Line~\ref{line:edgeprocessinginnerloop} of Algorithm~\ref{algo:edgprocessing}). All of the threads are
synchronized between iterations over separate subgraphs to avoid cache
interference. \edgeblocking improves the performance of some
algorithms by up to 2.94$\times$, as we show in
Table~\ref{tab:compedgeblocking}.

\algblockdefx[cfor]{CoopFor}{EndCoopFor}[1]
  {\textbf{coop for}~#1~\textbf{do}}

\makeatletter
\ifthenelse{\equal{\ALG@noend}{t}}%
  {\algtext*{EndCoopFor}}
  {}%
\makeatother

\makeatletter
\newcommand\fs@betterruled{%
  \def\@fs@cfont{\bfseries}\let\@fs@capt\floatc@ruled
  \def\@fs@pre{\vspace*{5pt}\hrule height.8pt depth0pt \kern2pt}%
  \def\@fs@post{\kern2pt\hrule\relax}%
  \def\@fs@mid{\kern2pt\hrule\kern2pt}%
  \let\@fs@iftopcapt\iftrue}
\floatstyle{betterruled}
\restylefloat{algorithm}
\makeatother

\renewcommand{\algorithmiccomment}[1]{\hfill$\triangleright$\textit{\color{gray}{#1}}}
\begin{algorithm}[!t]
\scriptsize
\caption{\small Algorithm for preprocessing the input graph when applying the \edgeblocking optimization. The algorithms takes as input a graph in the Coordinate (COO) format and the blocking size. The output is a new graph in COO format with the edges blocked.}
\label{algo:preprocessing}
\begin{algorithmic}[1]
\State \textbf{Input: } Number of vertices per segment $N$, Graph $G$ in COO format.
\State \textbf{Output: } Graph $\mathit{Gout}$ in COO format.
    \State $\mathit{numberOfSegments} \gets ceil(\mathit{G.num\_vertices / N})$
    \State $\mathit{segmentSize[numberOfSegments + 1]} \gets 0$
    \State $\mathit{Gout} \gets new\ \mathit{Graph}$
    \For{$e : \mathit{G.edges} $} \label{line:preprocessloop1}
        \State $\mathit{segment} \gets floor(\mathit{e.dst/N})$
	\State $\mathit{segmentSize[segment]} \gets \mathit{segmentSize[segment]} + 1$ \Comment{Count the number of edges in each subgraph.}
    \EndFor
    \State $\mathit{prefixSum} \gets \mathit{prefix\_sum(segmentSize)}$ \Comment{Identify the starting point for each subgraph using prefix sum.}
    \For{$e: \mathit{G.edges} $} \label{line:preprocessloop2}
        \State $\mathit{segment} \gets floor(\mathit{e.dst/N})$
        \State $\mathit{index} \gets \mathit{prefixSum[segment]}$
        \State $\mathit{Gout.edges[index]} \gets e$ \Comment{Add edge to the appropriate subgraph.}
        \State $\mathit{prefixSum[segment]} \gets \mathit{prefixSum[segment]} + 1$
    \EndFor
    \State $\mathit{Gout.numberOfSegments} \gets \mathit{numberOfSegments}$
    \State $\mathit{Gout.segmentStart} \gets \mathit{prefixSum}$
    \State $\mathit{Gout.N} \gets N$
    
\end{algorithmic}
\end{algorithm}

\begin{algorithm}[!t]
\scriptsize
\caption{\small Implementation of the \textit{edgeset.apply} operator with the \edgeblocking optimization. Here, the input graph is preprocessed and the edges are blocked. The function \textit{process\_edge} is applied to each edge in the graph. Note that the \edgeblocking optimization can be applied only when all the edges in the graph are being processed. }
\label{algo:edgprocessing}
\begin{algorithmic}[1]
\State \textbf{Input: } Graph $G$ in COO format. 

\For {$\mathit{segIdx}: \mathit{G.numberOfSegments}$} \label{line:edgeprocessingouterloop} \Comment{Iterate over the subgraphs.}
    \State $\mathit{start\_vertex} \gets \mathit{G.N} * \mathit{segIdx}$
    \State $\mathit{end\_vertex} \gets \mathit{G.N} * (\mathit{segIdx+1})$  
    \CoopFor {$\mathit{eid}\ in\ \mathit{G.segmentStart[segIdx-1]}:\mathit{G.segmentStart[segIdx]}$} \label{line:edgeprocessinginnerloop}
        \State $e \gets \mathit{G.edges[eid]}$
        \State $\mathit{process\_edge(e.src, e.dst)}$
    \EndCoopFor
    \State $\mathit{sync\_threads()}$ 
\EndFor
\end{algorithmic}
\end{algorithm}

\myparagraph{\TwceFullname (\twce)} Inspired by load balancing schemes in previous work~\cite{hong2017multigraph,Pai2016,davidson2014work,liu2015enterprise,KhorasaniGPU2015,merrill2012scalable}, \twce is a load balancing
strategy that further reduces the runtime overhead of \twc by
performing \twc style assignment within each CTA instead of across all
CTAs. Similar to \cm, each CTA starts with an equal number of vertices. Within each CTA, \twce partitions edges of the vertices into chunks that are processed by the entire CTA, a warp, or an individual thread. These partitions are processed in separate stages to minimize divergence.

Algorithm~\ref{algo:etwc} shows the \twce load balancing strategy. Each thread has a unique ID ($idx$), and 3 queues ($Q[0]$, $Q[1]$, $Q[2]$). These 3 queues correspond to the edges to be processed in the 3 stages. The outgoing edges of each vertex are partitioned and added to these queues such that each partition is an exact multiple of the chunk size starting from the greatest chunk size (CTA size) (Line 10--21). This ensures that fewer edges are processed in the individual thread stage, which is prone to divergence. 

For each stage, the representative thread (the first thread in the CTA or warp, or the thread itself) dequeues the 3-tuple from the corresponding queue and broadcasts it to the other threads in the CTA or warp (Line 24--26) for cooperatively processing the edges in a cyclic fashion (Line 27--29). These queues are managed in shared memory to reduce the overhead of the enqueue and dequeue operations.

\begin{algorithm}[!t]
\scriptsize
\caption{\small Pseudocode for the implementation of ETWC load balancing strategy. 
}
\algdef{SE}[SUBALG]{Indent}{EndIndent}{}{\algorithmicend\ }%
\algtext*{Indent}
\algtext*{EndIndent}

\label{algo:etwc}
\begin{algorithmic}[1]
\State \textbf{Input: } Graph $G$ in CSR format, input frontier ($\mathit{input\_frontier}$), and three queues $Q[0, 1, 2]$, and the number of threads to process edges of a vertex in three stages $Stage\_gran[0, 1, 2]$.

\State $\mathit{idx} = \mathit{threadblockID} * \mathit{threadblockSIZE} + \mathit{threadID}$
\State Initialize $\mathit{Q[0, 1, 2]}$

\If {$\mathit{idx} < \mathit{input\_frontier.size}$} 
    \State $\mathit{src\_id} \gets \mathit{getFrontierElement}(\mathit{input\_frontier}, \mathit{idx});$ 
    \State $\mathit{size} \gets \mathit{G.rowptr[src\_id+1]} - \mathit{G.rowptr[src\_id]}$ \Comment{The degree of a vertex.}
    \State $\mathit{start\_pos} \gets \mathit{G.rowptr[src\_id]}$ 
    \State $\mathit{end\_pos} \gets \mathit{G.rowptr[src\_id+1]}$ \Comment{State\_gran[2] is usually CTASize.} 
    \State $\mathit{Stage\_elt[2]} \gets \lfloor size / \mathit{Stage\_gran[2]} \rfloor \times \mathit{Stage\_gran[2]}$ 
    \If {$\mathit{Stage\_elt[2]} > 0$}
        \State $\mathit{Q[2].Enqueue}(\mathit{\{start\_pos, start\_pos + Stage\_elt[2], src\_id\}});$
	\State $\mathit{start\_pos} \gets \mathit{start\_pos} + \mathit{Stage\_elt[2]}$ 
	\State $\mathit{size} \gets \mathit{size} - \mathit{Stage\_elt[2]}$ 
    \EndIf
    \State $\mathit{Stage\_elt[1]} \gets \lfloor \mathit{size} / \mathit{Stage\_gran[1]} \rfloor \times \mathit{Stage\_gran[1]}$ \Comment{WarpSize.}
    \If {$\mathit{Stage\_elt[1]} > 0$}
        \State $\mathit{Q[1].Enqueue}(\mathit{\{start\_pos, start\_pos + Stage\_elt[1], src\_id\}});$
	\State $\mathit{start\_pos} \gets \mathit{start\_pos} + \mathit{Stage\_elt[1]}$ 
	\State $\mathit{size} \gets \mathit{size} - \mathit{Stage\_elt[1]}$ 

    \EndIf
    \If {$\mathit{size} > 0$}
        \State $\mathit{Q[0].Enqueue}(\{\mathit{start\_pos, end\_pos, src\_id\}});$
    \EndIf
\EndIf
\State $\mathit{sync\_threads}()$

\For{$i: 0,1,2 $}
    \While {!$\mathit{isEmpty.Q[i]}$}
        \If {$\mathit{threadID}\ \%\ \mathit{Stage\_gran[i]} == 0$} \Comment{Only representative thread.} 
            \State $\mathit{\{start\_pos, end\_pos, src\_id\}} \gets \mathit{Q[i].Dequeue()}$
            \State $Broadcast(\mathit{\{start\_pos, end\_pos, src\_id\}}, \mathit{Stage\_gran[i]})$
        \EndIf
        \CoopFor {$\mathit{eid}\ in\ \mathit{start\_pos:end\_pos-1}$} 
            \State $\mathit{dst\_id} \gets \mathit{G.edges[eid]}$
            \State $process\_edge(\mathit{src\_id}, \mathit{dst\_id})$ \Comment{Done by Stage\_gran[i] threads.}
        \EndCoopFor
    \EndWhile
\EndFor

\end{algorithmic}
\end{algorithm}

\subsection{CPU backend implementation}
\GG preserves the high-performance C++ CPU code generation backend
from \graphit. We changed the mid-end analyses and transformations in
such a way that they do not affect code generation for CPUs. This was
done to avoid compromising performance on CPUs. This is important because some applications like \deltastepping perform better on CPUs because of a limited amount of parallelism. We will compare the performance of CPUs vs GPUs for this application in Section~\ref{sec:evaluation}.

\subsection{Auto-tuning}
The \GG compiler exposes a large optimization space, with about
$10^6$ combinations of different schedules.  Even
without the hybrid schedules that involve two traversal directions,
the compiler can generate up to 288 combinations of schedules for each
direction (see Table~\ref{tab:opt_space}).  On top of that, integer
and floating-point parameters, such as the value of delta  for
Delta-stepping, blocking size of \edgeblocking, and thresholds for
hybrid schedules, need to be appropriately selected for each input
graph and algorithm. Searching through the huge optimization space
exhaustively is very time-consuming.

To navigate the schedule space more efficiently, we built an
auto-tuner for \GG using OpenTuner~\cite{ansel:pact:2014}.  For each
direction, the auto-tuner chooses among all 288 combinations of
options for load balancing, deduplication, output frontier strategy,
blocking, traversal direction, and kernel fusion.  For
direction-optimized schedules that involve two traversal directions,
the auto-tuner combines together two sets of schedules, one for each
direction.  The auto-tuner converges within 10 minutes on each input
graph for most algorithms and produces a schedule that matches the
performance of hand-optimized schedules.

%% file: evaluation.tex
\section{Evaluation}
\label{sec:evaluation}

In this section, we compare the performance of the code generated from
\GG's GPU backend with state-of-the-art GPU graph frameworks and
libraries on 5 graph algorithms and 9 different graph inputs. We also
study the performance tradeoffs and effectiveness of some key
optimizations. All of the GPU experiments are performed on an
NVIDIA Titan Xp (12 GB memory, 3MB L2 cache, and 30 SMs) and an NVIDIA
Volta V100 (32 GB memory, 4MB L2 cache, and 80 SMs). For the CPU performance evaluations,
we use a dual-socket Intel Xeon
E5-2695 v3 CPUs system with 12 cores per processor, for a total of 24
cores and 48 hyper-threads with 128 GB of DD3-1600 memory and 30 MB last level cache on each socket.

\myparagraph{Datasets} We list the input graphs used for our
evaluation in Table~\ref{tab:benchmarks}, along with their
sizes. These are the same datasets used to evaluate
\gunrock~\cite{wang2017gunrock}. 
Out of the 9 graphs, \gorkut, \gtwitter, \glivejournal, \gsinaweibo,
\ghollywood, and \gindochina have power-law degree distributions while
\gusad, \groadca, and \groadcentral are road graphs with bounded
degree distributions.

\setlength{\belowcaptionskip}{-0pt}
\begin{table}[t]
\vspace{2mm}
    \centering
    \begin{tabular}{l|rr}
        Graph Input & Vertex count & Edge count \\ \cline{1-3}
        \gorkut~\cite{nrvis} (\sgorkut) & 2,997,166 & 212,698,418 \\ 
        \gtwitter~\cite{nrvis} (\sgtwitter) & 21,297,772 & 530,051,090 \\ 
        \glivejournal~\cite{davis11acm-florida-sparse} (\sglivejournal) & 4,847,571 & 85,702,474 \\ 
        \gsinaweibo~\cite{nrvis} (\sgsinaweibo) & 58,655,849 & 522,642,066 \\ 
        \ghollywood~\cite{davis11acm-florida-sparse} (\sghollywood) & 1,139,905 & 112,751,422 \\ 
        \gindochina~\cite{davis11acm-florida-sparse} (\sgindochina)& 7,414,865 & 301,969,638 \\ 
        \gusad~\cite{road-graph} (\sgusad)& 23,947,347 & 57,708,624 \\ 
        \groadcentral~\cite{davis11acm-florida-sparse} (\sgroadcentral)& 14,081,816 & 33,866,826 \\ 
        \groadca~\cite{davis11acm-florida-sparse} (\sgroadca) & 1,971,281 & 5,533,214 \\ \cline{1-3}
    \end{tabular}
    \caption{Graph inputs used for evaluation. The edge count shows the number of undirected edges. 
    }
    \label{tab:benchmarks}
    
\end{table}

\myparagraph{Algorithms} We evaluate the performance of \GG and the
other frameworks on five algorithms: PageRank~(\pr), Breadth-First
Search~(\bfs), \deltastepping for Single-Source Shortest
Paths~(\sssp), Connected Components~(\cc) and Betweenness
Centrality~(\bc). These algorithms evaluate different aspects of the
compiler and give insights into how each optimization helps with
performance. \pr is a topology-driven algorithm that iterates over all
edges in every round and is useful in evaluating the performance
benefits of the \edgeblocking optimization. \bfs and \bc greatly
benefit from direction-optimization on graphs with a power-law
degree distribution. We use the hybrid scheduling API to get maximum
performance for these algorithms. \deltastepping makes use of the
priority queue API. Both \bfs and \deltastepping benefit from
\kernelfusion for high-diameter road graphs. \cc uses the algorithm by
Soman et al.~\cite{soman2010fast} and benefits from carefully choosing
the load balancing strategy.

\myparagraph{Comparison Frameworks} We compare \GG's performance with
three state-of-the-art GPU graph processing frameworks:
\gunrock~\cite{wang2017gunrock}, \gswitch~\cite{GSwitch2019}, and
\sepgraph~\cite{wang2019sep}. Both \gunrock and \gswitch have
optimized implementations of \bfs for power-law graphs using
direction-optimization. \sepgraph improves the performance of
\deltastepping and \bfs on high-diameter graphs by fusing kernel
launches across iterations like in our \kernelfusion optimization.

\gswitch chooses among several optimal parameters for traversal
direction, load balancing, and frontier creation using a learned
decision tree that makes use of graph characteristics and runtime
metrics. Each framework implements all of the algorithms that we evaluate,
except for \sepgraph, which does not implement \cc, \pr, and \bc. We did not compare \GG with Groute and IrGL because Groute is outperformed by SEP-Graph~\cite{wang2019sep}, and IrGL is not publicly available.
  
\begin{table}[t]
  \vspace{2mm}
  \centering
      \setlength{\tabcolsep}{5pt}
    \begin{tabular}{l|r|r|r|r|r}
    Framework           & \pr   & \bfs  & \deltastepping    & \cc       & \bc   \\ \cline{1-6}
    \gunrock            & 2207  & 2189  & 1438              & 3014 & 1792             \\ 
    \gswitch            & 159   & 164   & 203               & 160 & 280              \\ 
    \sepgraph           & --   & 481   & 473               & -- & --        \\ 
    \GG    (algorithm+schedule) & \textbf{61}    & \textbf{66}    & \textbf{50}                & \textbf{62} & \textbf{128}        \\ 
    \end{tabular}
    \caption{Number of lines of code for the five algorithms written using \gunrock, \gswitch, \sepgraph, and \GG. \sepgraph does not implement \cc \bc and \pr. The shortest code length for each algorithm is shown in \textbf{bold}.}
    \label{tab:linecount}
\end{table}

\newcommand*{\lastc}[1]{\multicolumn{1}{r|[1.2pt]}{#1}}
\newcommand*{\nlastc}[1]{\multicolumn{1}{r|}{#1}}

\begin{table*}
\vspace{2mm}
\centering
	\begin{tabu}[width=\textwidth]{|[1.2pt]p{0.55cm}|[1.2pt]R{1.0cm}|R{1.1cm}|R{1.0cm}|R{1.0cm}|[1.2pt]R{1.1cm}|R{1.1cm}|R{1.0cm}|R{1.0cm}|[1.2pt]R{1.0cm}|[1.2pt]}
		\tabucline[1.2pt]{-}
				 & \multicolumn{3}{c|[1.2pt]}{\pr (time per round)} & \multicolumn{3}{c|[1.2pt]}{\cc} & \multicolumn{3}{c|[1.2pt]}{\bc} \\ \tabucline[1.2pt]{-}
		Graph		 & \GG	& \sgunrock & \lastc{\sgswitch} & \nlastc{\GG} & \sgunrock & \lastc{\sgswitch} & \GG & \nlastc{\sgunrock} & \sgswitch \\ \hline	
		\sgorkut         & \textbf{14.18}     & 60.63     & \lastc{117.10}   & \nlastc{\textbf{63.31}}     & 71.95     & \lastc{76.85}    & \textbf{26.22} 	& \nlastc{213.93}	& 37.93	  \\ \hline   
		\sgtwitter       & \textbf{77.86}     & 113.55    & \lastc{211.03}   & \nlastc{\textbf{196.78}}    & 374.59    & \lastc{OOM}      & 174.83	& \nlastc{505.31}	& \textbf{122.81}  \\ \hline
		\sglivejournal   & \textbf{7.68}      & 17.67     & \lastc{OOM}      & \nlastc{\textbf{24.65}}     & 35.96     & \lastc{27.81}    & 28.93	& \nlastc{88.04}		& \textbf{26.98}  \\ \hline	
		\sgsinaweibo     & \textbf{102.11}    & 178.70    & \lastc{338.79}   & \nlastc{\textbf{276.04}}    & 439.51    & \lastc{OOM}      & \textbf{204.32}	& \nlastc{1095.60}	& 415.06  \\ \hline	
		\sghollywood     & \textbf{7.01}      & 22.67     & \lastc{OOM}      & \nlastc{\textbf{12.04}}     & 37.43     & \lastc{18.23}    & \textbf{12.20}	& \nlastc{29.03}		& 79.44	  \\ \hline	
		\sgindochina     & 18.24     & 13.16     & \lastc{\textbf{9.30}}     & \nlastc{\textbf{31.66}}     & 235.92    & \lastc{43.10}    & 35.78	& \nlastc{47.97}		& \textbf{10.70}  \\ \hline	
		\sgusad          & \textbf{6.32}      & 10.53     & \lastc{7.62}     & \nlastc{\textbf{20.66}}     & 74.45     & \lastc{31.21}    & \textbf{302.22}	& \nlastc{987.93}	& 564.53  \\ \hline	
		\sgroadcentral   & \textbf{5.56}      & 9.96      & \lastc{8.86}     & \nlastc{\textbf{27.03}}     & 48.22     & \lastc{27.13}    & \textbf{239.55}	& \nlastc{632.97}	& 332.91  \\ \hline	
		\sgroadca        & \textbf{0.43}      & 0.94      & \lastc{0.47}     & \nlastc{\textbf{1.72}}      & 5.82      & \lastc{3.04}     & \textbf{24.05}	& \nlastc{86.44}		& 39.51	  \\ \tabucline[1.2pt]{-}
			& \multicolumn{4}{c|[1.2pt]}{\sssp with \deltastepping} & \multicolumn{4}{c|[1.2pt]}{\bfs} & \\ \tabucline[1.2pt]{-}
		Graph	& \GG & \sgunrock & \sgswitch & SEP-G & \GG & \sgunrock & \sgswitch & SEP-G & \hyp{} \\ \hline
		
		\sgorkut         & \textbf{94.30}     & 978.44            & 550.38            & 434.77     & \textbf{1.75}     & 1.92      & 1.94      & 6.40      & \\ \hline    
		\sgtwitter       & \textbf{114.34}    & 264.54            & 233.54            & 237.12     & \textbf{21.13}     & OOM    & 22.52     & 40.18       & \\ \hline
		\sglivejournal   & \textbf{54.46}     & 260.19            & 172.99            & 172.07     & 4.66      & 4.80    & \textbf{3.68}      & 11.63       & \\ \hline
		\sgsinaweibo     & \textbf{685.66}    & 3470.59           & 1933.29           & 2296.01     & 20.17     & OOM    & \textbf{18.96}     & 93.69       & \\ \hline
		\sghollywood     & \textbf{18.44}     & 74.90             & 47.54             & 92.92      & \textbf{1.89}      & 2.04      & 2.30      & 4.76     & \\ \hline
		\sgindochina     & \textbf{120.42}    & 232.55            & 268.95            & 511.70     & \textbf{10.68}     & 15.92     & 70.69     & 55.15    & \\ \hline
		\sgusad          & 601.08     & 53518.87          & 1238.03            & \textbf{440.21}     & \textbf{73.15}     & 442.19    & 119.93    & 84.29    & \\ \hline
		\sgroadcentral   & 337.62    & 29443.51          & 642.38            & \textbf{286.03}     & \textbf{49.59}     & OOM    & 84.98     & 61.51       & \\ \hline
		\sgroadca        & 17.01     & 89.18             & 27.80             & \textbf{16.48}      & \textbf{6.13}      & OOM    & 10.29     & 6.88        & \\ \tabucline[1.2pt]{-}

	\end{tabu}	
    \caption{
    Execution time in milliseconds for the \totalalgo algorithms on \totalgraph input graphs for the 4 frameworks in comparison, \GG, \gunrock (GU), \gswitch (GSW), and \sepgraph (SEP-G) running on an NVIDIA Titan Xp GPU. The fastest result per algorithm-graph combination is shown in \textbf{bold}.
	    The \pr, \bfs, \bc, and \cc algorithms use unweighted and symmetrized graphs. Single-Source Shortest Path (SSSP) with \deltastepping uses the original graphs (not symmetrized) with edge weights.
	    Uniformly random integer weights between 1--1000 are added for \gorkut, \gsinaweibo, \ghollywood, and \gindochina because they did not originally have edge weights. OOM indicates that the framework ran out of memory for the particular input. 
    }
    \label{tab:mastereval}
	\begin{tabu}[width=\textwidth]{|[1.2pt]p{0.55cm}|[1.2pt]R{1.0cm}|R{1.1cm}|R{1.0cm}|R{1.0cm}|R{1.1cm}|R{1.1cm}|R{1.0cm}|R{1.0cm}|R{1.0cm}|[1.2pt]}
		\tabucline[1.2pt]{-}
				 & \multicolumn{3}{c|[1.2pt]}{\pr (time per round)} & \multicolumn{3}{c|[1.2pt]}{\cc} & \multicolumn{3}{c|[1.2pt]}{\bc} \\ \tabucline[1.2pt]{-}
		Graph		 & \GG	& \sgunrock & \lastc{\sgswitch} & \nlastc{\GG} & \sgunrock & \lastc{\sgswitch} & \GG & \nlastc{\sgunrock} & \sgswitch \\ \hline	
		\sgorkut         & \textbf{7.9}     & 20.59     & \lastc{32.77}  &  \textbf{15.18}     & \hyp{}        & \lastc{16.33}   & \textbf{20.05} 	& 213.93	& \lastc{20.56}		 \\ \hline    
		\sgtwitter       & \textbf{39.44}     & 43.04     & \lastc{103.88} &  \textbf{134.95}    & \hyp{}        & \lastc{137.12}  & 90.69	& 505.31	& \lastc{\textbf{56.44}}	 \\ \hline	
		\sglivejournal   & \textbf{3.76}      & 7.66      & \lastc{6.49}   &  \textbf{7.9}      & \hyp{}        & \lastc{11.12}   & 21.31	& 88.04		& \lastc{\textbf{16.88}}	 \\ \hline
		\sgsinaweibo     & \textbf{50.18}     & 66.41     & \lastc{163.88} &  \textbf{173.59}    & \hyp{}        & \lastc{290.02}  & \textbf{154.06}	& 1095.60	& \lastc{216.05}	 \\ \hline
		\sghollywood     & \textbf{4.27}      & 6.47      & \lastc{8.67}   &  7.4      & \hyp{}        & \lastc{\textbf{6.85}}    & 11.26	& 29.03		& \lastc{\textbf{6.08}}		 \\ \hline
		\sgindochina     & \textbf{13.99}     & 16.64     & \lastc{46.55}  &  22.31     & \hyp{}        & \lastc{\textbf{4.46}}    & \textbf{30.09}	& 47.97		& \lastc{55.58}		 \\ \hline
		\sgusad          & 3.1      & \textbf{2.68}      & \lastc{3.22}   &  \textbf{12.44}     & \hyp{}        & \lastc{17.94}   & \textbf{330.82}	& 987.93	& \lastc{536.95}	 \\ \hline
		\sgroadcentral   & \textbf{2.69}      & 2.61      & \lastc{2.97}   &  \textbf{10.01}     & \hyp{}        & \lastc{14.14}   & \textbf{213.39}	& 632.97	& \lastc{336.78}	 \\ \hline
		\sgroadca        & \textbf{0.22}      & 0.27      & \lastc{0.29}   &  \textbf{1.34}      & \hyp{}        & \lastc{2.79}    & \textbf{32.79}	& 86.44		& \lastc{47.06}	         \\ \tabucline[1.2pt]{-}

				 & \multicolumn{4}{c|[1.2pt]}{\sssp with \deltastepping} & \multicolumn{4}{c|[1.2pt]}{\bfs} & \\ \tabucline[1.2pt]{-}
		Graph		 & \GG & \sgunrock & \sgswitch & \lastc{SEP-G} & \GG & \sgunrock & \sgswitch & \lastc{SEP-G} & \hyp{} \\ \hline
		\sgorkut         & \textbf{50.09}     & 303.15        & 199.59            & \lastc{164.69}   & \textbf{1.51}      & 1.66    & \textbf{1.51}     & \lastc{5.72}  & \\ \hline 
		\sgtwitter       & \textbf{61.85}      & 101.33         & 132.94            & \lastc{117.97}   & 14.73     & 13.52   & \textbf{10.60}    		& \lastc{38.18} & \\ \hline    	
		\sglivejournal   & \textbf{40.05}     & 131.91         & 77.95             & \lastc{103.40}   & \textbf{2.68}      & 3.63    & 3.05     		& \lastc{9.59}  & \\ \hline    	
		\sgsinaweibo     & \textbf{375.42}    & 1686.39        & 1062.56           & \lastc{1066.57}  & 13.86     & 94.77   & \textbf{12.26}    		& \lastc{70.70} & \\ \hline    	
		\sghollywood     & \textbf{18.04}     & 37.42          & 26.77             & \lastc{51.75}    & 1.65      & 1.67    & \textbf{1.60}     		& \lastc{4.81}  & \\ \hline    	
		\sgindochina     & 100.49    & \textbf{25.58}         & 211.85            & \lastc{350.58}   & \textbf{9.03}     & 12.57   & 40.60    		& \lastc{39.39} & \\ \hline    	
		\sgusad          & 253.11    & 788.23        & 390.23            & \lastc{\textbf{191.08}}   & \textbf{74.26}     & 775.16  & 186.43   		& \lastc{97.62} & \\ \hline    	
		\sgroadcentral   & 168.7    & 429.24        & 222.05            & \lastc{\textbf{128.02}}   & 118.52    & 434.01  & \textbf{115.13}   		& \lastc{65.49} & \\ \hline    	
		\sgroadca        & 21.08     & 66.79         & 32.47             & \lastc{\textbf{19.46}}    & 10.09     & 68.35   & 16.76    		& \lastc{\textbf{9.33}} & \\ \tabucline[1.2pt]{-}

	\end{tabu}
	    \caption{
    Execution time in milliseconds for the same experiments and comparison frameworks in Table~\ref{tab:mastereval} running on an NVIDIA V100 GPU. The fastest result per algorithm-graph combination is shown in \textbf{bold}. \gunrock's \cc implementation did not work correctly on this GPU. }
    \label{tab:masterevalvolta}
\vspace{-2em}
\end{table*}

\subsection{Comparison with other Frameworks} 
Tables~\ref{tab:mastereval} and~\ref{tab:masterevalvolta} show the
execution times of all of the algorithms in \GG and the other
frameworks on the Titan Xp and V100 machines, respectively. \GG
outperforms the next fastest of the three frameworks on \fastercount
out of \totalexpr experiments by up to \uptospeedup. Table~\ref{tab:linecount} shows the number of lines of code for each algorithm in each framework. \GG always uses
significantly fewer lines of code compared to other frameworks.  

\myparagraph{\pr} \GG has the fastest \pr on 16 out of 18
experiments. Compared to \gunrock and \gswitch, \GG is up to 4.2x faster. This is mainly
because of the \edgeblocking optimization that reduces the number of
L2 cache misses, as described in
Section~\ref{sec:new-optimizations}. The results are even better on the V100 GPU because of its
larger L2 cache size. The graph is divided into fewer subgraphs
reducing the iteration overhead.

\myparagraph{\bfs} \GG has the fastest \bfs on 11 of the 18 experiments. \GG
outperforms \gswitch and \gunrock by up to 6.04x and 1.63x,
respectively, on the road graphs because \gunrock and \gswitch do not
use the \kernelfusion optimization to reduce kernel launch
overheads as discussed in Section~\ref{sec:gpu_backend_impl}.
\sepgraph is only up to 1.24x slower than \GG on the road graphs
because it uses asynchronous execution, which is beneficial
  for high-diameter graphs. However, the better load balancing
achieved by \twce makes \GG faster than \sepgraph.

On the power-law graphs, direction-optimization is very
effective. Both \gunrock and \gswitch use direction-optimization and
hence the performance of \GG is very close to both of
them. \gswitch and \gunrock also use idempotent
label updates, which introduces a benign race instead of using expensive compare-and-swap. This optimization is specific for \bfs and is hard to generalize in a compiler like \GG. Even on road graphs, \GG gains significant speedups over
\gswitch and \gunrock because of the \kernelfusion optimization as the
kernel launch overhead is still the dominating factor. 

The \gindochina graph is a
special case of a power-law graph that does not benefit from
direction-optimization because the number of active vertices never crosses our threshold for \stt{PULL} strategy to be beneficial over \stt{PUSH} strategy.
Both \gunrock and \gswitch use direction-optimization for \gindochina
and suffer from the extra work done in the \stt{PULL}
direction. \GG uses the more efficient \stt{PUSH}-only
 schedule. 

\myparagraph{\cc} For \cc, \GG is the fastest on 16 out of the 18 experiments. \GG is up to 3.4x faster than the next fastest framework. All
of the frameworks use the same algorithm and execution strategies. We
tune the performance by choosing different load balancing strategies
for each graph (\twce for power-law graphs and \cm for
road graphs). The speedups are not as significant on the Volta GPU as
compared to the Pascal GPU, because Volta generation GPUs have
semi-warp execution that reduces the importance of load balancing
techniques.

\myparagraph{\deltastepping} \GG has the fastest \deltastepping
performance on 11 of the 18 graph inputs and runs up to 4.61x faster
than the next fastest framework.  \deltastepping needs the
\kernelfusion optimization for road graphs because of their high
diameter and the low number of vertices processed in each iteration.
\gswitch and \gunrock lack this optimization and thus are slower on road
graphs by up to 2.05x and 89x, respectively.  On power-law graphs, \GG
benefits from the better \twce load balancing strategy and performs up
to 5.11x faster than the next fastest framework.  On road-graphs,
\sepgraph executes up to 1.36x times faster because of the highly
optimized asynchronous execution. Finally, in some cases, \GG runs more slowly on Volta than on
Pascal because the kernel launch overhead and synchronization costs
are higher on Volta than on Pascal due to its larger number of SMs
(\deltastepping does not have enough parallelism to utilize all of the
SMs). The other frameworks also have a similar slowdown when moving from Pascal to Volta. Table~\ref{tab:cpu_gpu_comp} shows that on road-graphs, \GG's CPU implementation is 2x faster than \sepgraph's. This highlights the need for generating code for both CPU and GPU from the same high-level representation.

\myparagraph{\bc} \GG has the fastest \bc performance on 12 out of 18 experiments.
\GG runs up to 2.03x faster than the next fastest
framework. \gunrock does not use
direction-optimization for \bc, which is critical for high performance
on power-law graphs.  
\gswitch uses direction-optimization, but \GG outperforms \gswitch
because of the better load balancing from \twce. For high-diameter
graphs, \GG benefits greatly from the \kernelfusion optimization. Both
\gswitch and \gunrock do not implement \kernelfusion. 

\subsection{Comparison against CPU}
\label{sec:cpucomp}
We also compare the performance of \GG-generated CPU implementations
with \GG-generated GPU
implementations running on the Titan Xp Pascal generation GPU. \GG generates the
same CPU implementation as \graphit and
\OG~\cite{zhang2019prioritygraph,graphit:2018}. 
On \pr, \bfs, and
\cc, the GPU implementations are faster because these algorithms can
easily utilize the large amount of parallelization and memory bandwidth available on the GPUs. On the other hand, \deltastepping,
which has less parallelism available when running on road graphs,
executes up to 2.07$\times$ faster on the CPU due to more powerful
cores and larger caches. The execution times for \deltastepping on the GPU
and CPU can be found in Table~\ref{tab:cpu_gpu_comp}.  Another advantage of
CPUs is that they can process graphs that are much larger than the GPU memory more
efficiently. These experiments provide
evidence for our claim that neither CPUs nor GPUs are suitable for all algorithms and inputs and shows how a framework that can generate code for both CPUs and GPUs from the same input can help achieve high performance. 

\subsection{Performance of \twce and \edgeblocking}
In this section, we compare the performance of the two optimizations that we introduce in \GG, \twce and \edgeblocking.  To
evaluate the performance of the \twce load balancing scheme, we run
the \bfs algorithm on all nine graph inputs. We fix the direction to
\stt{PUSH} and do not use kernel fusion. 

\begin{table}[t]
   \vspace{2mm}
    \centering
    \begin{tabular}{l|r|r|r}
    Delta-Stepping for SSSP      & \GG CPU                   & \GG GPU           & \sepgraph     \\ \cline{1-4}
    \gusad                      & \textbf{212.30}           & 601.80            & 440.21        \\ 
    \groadcentral               & \textbf{162.49}           & 337.62            & 286.03        \\ 
     \gorkut                    & 106                       & \textbf{94.3}     & 434.77        \\ 
     \glivejournal              & 90.05                     & \textbf{54.46}    & 172.07        \\ \cline{1-4}
    \end{tabular}
    \caption{Comparisons of \GG-generated CPU implementations, \GG-generated GPU implementations, and \sepgraph on SSSP with \deltastepping. The GPU experiments are run on the Pascal generation GPU. The running times are in milliseconds. The fastest result per graph is shown in \textbf{bold}. We did not count the data transfer time from CPU to GPU. We do not show \gunrock and \gswitch because they are always outperformed by \sepgraph for \sssp.}
    \label{tab:cpu_gpu_comp}
\end{table}

We only choose among the \twce, \twc, and \cm load balancing schemes (\twc and
\cm are the best performing on social and road graphs, respectively,
 and hence other load balancing schemes are not being
compared).  The results
are shown in Table~\ref{tab:comptwce}. \cm is faster than \twc
on graphs that have a regular degree distribution, like road graphs
and the \gindochina graph, while \twc performs better on social graphs
with power-law degree
distributions because it is able to separate out vertices of different
degrees. \twc is slower on road graphs because of the overhead from
load balancing. \twce outperforms the other two load balancing schemes
on 7 of the 9 graphs. \twce does well both on social and road graphs
because it is able to effectively load balance vertices with varying
degrees without incurring a large overhead. This is because unlike
\twc, \twce balances the edges only locally within a CTA, thus
communicating only using shared memory. 
\begin{table}[t]
\vspace{-1em}
    \centering
    \begin{tabular}{l|r|r|r}
        Graph           & \twce     & TWC       & CM        \\ \cline{1-4}
        \gorkut         & 43.58     & \textbf{40.69}     & 42.24     \\ 
        \gtwitter       & \textbf{106.11}    & 107.57    & 116.06    \\ 
        \glivejournal   & 19.72     & 20.03     & \textbf{18.42}     \\ 
        \gsinaweibo     & \textbf{226.35}    & 230.00    & 230.03    \\ 
        \ghollywood     & \textbf{4.94}      & 5.79      & 8.17      \\ 
        \gindochina     & \textbf{11.38}     & 11.5      & 22.16     \\ 
        \gusad          & \textbf{136.64}    & 255.89    & 168.9     \\ 
        \groadcentral   & \textbf{91.2}      & 162.54    & 109.89    \\ 
        \groadca        & \textbf{13.1}      & 25.77     & 16.25     \\ \cline{1-4}
    \end{tabular}
    \caption{Execution time (in milliseconds) of \bfs (\tablelst{PUSH} only) using \twce, \twc, and \cm load balancing strategies on the Titan Xp GPU. The fastest result per graph  is shown in \textbf{bold}.}
    \label{tab:comptwce}
\end{table}

\begin{table}[t]
    \vspace{2mm}
    \centering
    \begin{tabular}{l|r|r|r|r}
        Graph           & Without EB    & With EB   & Speedup & Preprocessing   	  \\ 
                        & 		&	    &         & time 			\\ \cline{1-5}
        \gorkut         & 41.75         & \textbf{14.18}     & 2.94x        & 22.75 	     \\ 
        \gtwitter       & 88.25         & \textbf{77.86}     & 1.13x        & 129.43 	     \\ 
        \glivejournal   & 15.67         & \textbf{7.68}      & 2.04x       & 20.65 	     \\ 
        \gsinaweibo     & 144.88        & \textbf{102.11}    & 1.41x        & 141.86 	     \\ 
        \ghollywood     & \textbf{7.01}          & 7.02      & 0.99x       & 12.00 	     \\ 
        \gindochina     & \textbf{18.24}         & 19.55     & 0.93x        & 32.25 	     \\ 
        \gusad          & 170.75        & \textbf{126.41}    & 1.35x 	   & 11.77 	     \\ 
        \groadcentral   & 167.82        & \textbf{111.37}    & 1.50x 	   & 8.20 	     \\ 
        \groadca        & 8.93          & \textbf{8.74}      & 1.02x 	   & 1.08 	     \\ \cline{1-5}
    
    \end{tabular}
    \caption{Execution time and preprocessing time (in milliseconds) per round of \pr with and without \edgeblocking on the Titan Xp GPU. The speedup column shows improvement in execution time with \edgeblocking enabled.}
    \label{tab:compedgeblocking}
\end{table}

Similarly, we also evaluate the performance gains of using
\edgeblocking by running \pr on all of the input graphs. We fix the
schedule to use edge-only load balancing (which is the fastest for
\pr) and compare the execution times with and without
\edgeblocking. The vertex data of these vertices fit in the L2 cache
of the GPU. Table~\ref{tab:compedgeblocking} shows the execution time
of \pr with \edgeblocking disabled and enabled and the speedup from
enabling it. We see that with \edgeblocking enabled, \pr runs up to
2.94x faster. However, \edgeblocking causes some slowdown on
\gindochina and \ghollywood because of degradation in work-efficiency and the graphs being already somewhat clustered (the IDs of the neighbors of each vertex are in a small range).
\edgeblocking requires us to preprocess the input graphs. Table~\ref{tab:compedgeblocking} shows that the overhead for preprocessing is less than the time required for two rounds of \pr and thus can be easily amortized across multiple runs. 

\begin{table}[t]
    \centering
    \begin{tabular}{l|r|r|r}
        Graph           & Without       & With   	     	& Speedup      		\\ 
                        & \kernelfusion	& \kernelfusion	     	&         	    	\\ \cline{1-4}
        \gorkut         & \textbf{1.51} 		&  2.21 		& 0.68x         		\\ 
        \gtwitter       & \textbf{14.73} 	&  21.42 		& 0.69x        		\\ 
        \glivejournal   & \textbf{2.68}		&  5.24 		& 0.51x        		\\ 
        \gsinaweibo     & \textbf{13.86}		&  18.56 		& 0.74x        		\\ 
        \ghollywood     & \textbf{1.65} 		&  2.51 		& 0.65x        		\\ 
        \gindochina     & \textbf{9.03} 		&  14.26 		& 0.63x        		\\ 
        \gusad          & 261.28 	&  \textbf{74.26} 		& 3.51x        		\\ 
        \groadcentral   & 161.2 	&  \textbf{118.52} 		& 1.36x        		\\ 
        \groadca        & 22.85 	&  \textbf{10.09} 		& 2.26x        		\\ \cline{1-4}
    
    \end{tabular}
    \caption{Execution time (in milliseconds) for \bfs with and without \kernelfusion. The last column shows the speedup with \kernelfusion enabled on the V100 GPU.}
    \label{tab:kernelfusion_exp}
\end{table}

Finally, we evaluate the performance impact of  \kernelfusion. Although \kernelfusion is not a new technique presented in this paper, a novelty of the compiler is the algorithm for generating fused kernels for an arbitrary sequence of operations. Table~\ref{tab:kernelfusion_exp} shows the execution time of the \bfs algorithm with and without \kernelfusion on all of the graph inputs on the V100 GPU. We can clearly see that enabling \kernelfusion offers up to 3.51x speedup for road graphs, which have a large diameter and run for iterations with very few vertices to process per iteration. On the other hand, enabling \kernelfusion slows down the execution for power-law degree graphs by up to 2x because it significantly affects load balancing, which is critical for power-law degree graphs. Fused kernels also increase register pressure, which affects kernels that process a lot of vertices per iteration. These experiments show that it is critical to tune the kernel launch decision using the scheduling language.

%% file: conclusion.tex
\section{Conclusion}
\label{sec:conclusion}

We introduce the \GGfull for
writing high-performance graph algorithms for both CPUs and GPUs. \GG
also introduces a novel GPU scheduling language and allows users to
search through many different combinations of \factorsbroad
optimizations.  We propose two performance optimizations, \twce
and \edgeblocking, to improve load balancing and locality of edge
processing, respectively.  We evaluate \GG on \totalalgo algorithms
and \totalgraph graphs and show that it achieves up to \uptospeedup
speedup over the next fastest framework, and is the
fastest on \fastercount out of the \totalexpr experiments.

%% file: acks.tex
\section{Acknowledgments}
This research was supported by the MIT Research Support Committee Award, DARPA SDH Award \#HR0011-18-3-0007, Applications Driving Architectures (ADA) Research Center, a JUMP Center co-sponsored by SRC and DARPA, DARPA D3M Award \#FA8750-17-2-0126, DOE Early Career Award \#DE-SC0018947, NSF
CAREER Award \#CCF-1845763, and Google Faculty Research Award.